# Hydrogen storage in nanocrystalline high entropy material

Yogesh Kumar Yadav[1], Mohammad Abu Shaz[1], and Thakur Prasad Yadav[1, 2]
[1]Department of Physics, Institute of Science, Banaras Hindu University Varanasi-221005, India
[2]Department of Physics, Faculty of Science, University of Allahabad, Prayagraj-211002, India

**Abstract:**
In this study, a single-phase nanocrystalline Al-Cu-Fe-Ni-Cr high-entropy alloy (HEA) has been synthesized by mechanical alloying and comprehensively investigated for hydrogen storage responses evaluated in details. High-energy attritor ball mill was used to synthesize the alloy from elemental powder, and hexane medium was used as a process control agent. As synthesized materials was nanocrystalline in nature after 40 h of milling with a lattice parameter of 0.289 nm body-centered cubic (BCC) phase. As synthesized nanocrystalline Al-Cu-Fe-Ni-Cr HEA demonstrated remarkable hydrogen storage properties, absorbing 2.1 wt.% of hydrogen in 3 minutes at 300°C with 50 atm of hydrogen pressure. At the same temperature, it also desorbed about 1.6 wt.% of hydrogen in 6 minutes. These quick rates of absorption and desorption demonstrate how well the alloy absorbs and releases hydrogen. Additionally, the alloy showed outstanding cyclic stability, retaining almost all of its hydrogen capacity across 25 cycles with only a slight 0.2 wt.% loss. The nanocrystalline Al-Cu-Fe-Ni-Cr HEA is a potential option for hydrogen storage applications due to its outstanding cycle stability and fast kinetics of hydrogen storage and release.

## 1. Introduction:

According to the Sixth Assessment Report (AR6) of the Intergovernmental Panel on Climate Change (IPCC), human-induced global warming has caused previously unheard-of alterations in the Earth climate, with current readings of 1.1 degrees Celsius [1-2]. There are already a number of concerning trends [1,2]: The last decade was warmer than any time in the previous 125,000 years, CO2 concentrations are at levels not seen in at least two million years, and glacier retreat is occurring at its fastest rate in over 2,000 years. Ocean temperatures are rising faster than they have since the end of the last ice age, and sea levels are rising faster than they have in the last 3,000 years. The oceans are also becoming more acidic than they have been in 26,000 years. Climate change has a substantial impact on human populations and ecosystems, and the hazards will increase with every degree of warming [3]. The use of fossil fuels (coal, oil, and gas) is the main cause of this warming [4-5]. The use of fossil fuels, a primary cause of the greenhouse effect, has caused $CO_2$ concentrations to climb sharply, from 11 billion tonnes in 1960 to a projected 36.6 billion tonnes in 2023 [6]. We desperately need pollution-free, renewable alternative fuels to meet these issues. Because of its abundance and reputation as a clean, adaptable, and sustainable energy source, hydrogen energy becomes an essential solution [7]. Despite its exceptional energy density and carbon neutrality, hydrogen strong flammability makes storage and transportation difficult. In contrast, solid-state storage techniques are less expensive and safer than storing hydrogen in high-pressure cylinders or as a cryogenic liquid [8].

Despite the impressive hydrogen storage capacities demonstrated by metal hydrides, complex hydrides, metal-organic frameworks (MOFs), and carbon nanostructures, a number of challenges remain, including difficult synthesis processes, limited reversibility, and poor cyclability [9]. In light of these issues, utilising metal hydrides in conjunction with specific metals and alloys to store hydrogen offers a potential solution, potentially leading to high-capacity, low-cost storage that is more cyclable and reliable in terms of safety [10]. In this regard, HEAs, a unique class of metallic materials, have drawn interest for use in hydrogen storage applications. The HEAs are composed of five or more principal constituents in roughly similar quantities [11]. The high entropy effect, sluggish diffusion effect, severe lattice distortion impact, and cocktail effect are the four main effects linked to the distinctive characteristics of HEAs [12]. The development of efficient hydrogen storage materials depends heavily on these factors. Significant empty space is created for the storage of hydrogen as a result of the high entropy effect, which encourages the creation of a solid solution phase [13]. Grain formation is facilitated by the slow diffusion effect, which improves the kinetics of hydrogen absorption and desorption [14]. Larger interstitial sites are produced by the extreme lattice distortion effect, which enhances hydrogen storage capacity [15]. Lastly, the physical and chemical characteristics of HEAs are influenced by the cocktail effect, which results from the combination of these three factors [10].

The idea of hydrogen storage in HEAs was first presented in a groundbreaking paper by Kao et al. in 2010, which paved the way for further investigation into alloy composition optimisation, especially with regard to alloys that efficiently store hydrogen at ambient temperature. The alloy Co-Fe-Mn-Ti-V-Zr [16] was noted. A C14 Laves phase structure, which is advantageous for hydrogen storage because of its effective hydrogen absorption capacity, was demonstrated in one of the first HEA hydrogen storage successes. At room temperature, the reversible hydrogen capacity of this alloy was up to 1.6 wt.%. A hexagonal structure recognised for providing sites conducive to hydrogen occupancy, the C14 Laves phase is frequently seen in HEAs intended for hydrogen storage. The Ti-Zr-Cr-Mn-Fe-Ni alloy is another noteworthy HEA designed especially for hydrogen storage; it required no activation to reach 1.7 wt.% capacity at normal temperature [17]. This alloy quick kinetics of absorption and desorption, primarily in the C14 phase, allowed for effective hydrogen uptake and release. In their investigation of a $Ti_{20}Zr_{20}V_{20}Cr_{20}Ni_{20}$ HEA, Kumar et al. (2022) obtained a low plateau pressure below five atmospheres and 1.52 wt.% hydrogen storage [18]. Stability and ease of activation were proved by its single-phase C14 Laves structure, which are crucial characteristics for real-world applications. Another HEA, a hexanary HEA with the formula $Ti_{0.24}V_{0.17}Zr_{0.17}Mn_{0.17}Co_{0.17}Fe_{0.08}$ [19], was synthesised by Kumar et al. in 2023. With the help of its C14-type Laves phase, this alloy was able to quickly absorb 0.53 wt.% of hydrogen in 15 seconds and reach a maximum capacity of 0.72 wt.% after 150 minutes. The alloy favourable thermodynamics for hydrogen absorption and release were highlighted by the computed enthalpy values for hydrogen absorption and desorption, which were roughly -19.06 kJ/mol and -34.10 kJ/mol, respectively. Dangwal et al. (2024) advanced the field further by customising a HEA through careful element selection based on factors such as atomic-size mismatch and valence electron concentration [20]. By mixing the C14 Laves and BCC phases, their $TiV_2ZrCrMnFeNi$ alloy was able to store 1.6 wt.% of hydrogen at room temperature without activation. Rapid hydrogen absorption and desorption were made possible by the dual-phase composition, which benefited from the interphase boundaries between the C14 and BCC structures, which encouraged advantageous activation features. This alloy demonstrated effective and completely reversible hydrogen storage properties, underscoring the contribution of mixed-phase HEAs to improved hydrogen storage efficiency. Because of HEAs structural and compositional versatility, scientists have been able to synthesize alloys that optimise hydrogen uptake. Li et al. (2018), for instance, investigated a $(TiZr_{0.1})_xCr_{1.7-y}Fe_yMn_{0.3}$ series and discovered that hydrogen storage was enhanced by lowering the Fe content; the (TiZr$_{0.1}$)$_{1.1}$Cr$_{1.5}$Fe$_{0.2}$Mn$_{0.3}$ variety achieved 2.0 wt.% capacity at 274 K [21]. A hydrogen dissociation enthalpy of 24.3 kJ/mol demonstrated the alloy favourable thermodynamic stability and reversible hydrogen capacity of 1.65 wt.%. The lower Fe content was thought to have achieved the best balance between stability and capacity by modifying plateau pressures for efficient hydrogenation and dehydrogenation. An AB-type Ti-Zr-Nb-Cr-Fe-Ni HEA synthesised by arc melting was studied by Andrade et al. (2023) and showed two different C14 Laves phases with minor differences in unit cell dimensions [22]. Using pressure-composition-temperature (PCT) isotherms, they discovered that the alloy could absorb 1.5 wt.% hydrogen at room temperature without the need for activation procedures. Complete desorption, however, was not achievable in these circumstances. The alloy exhibited 1.1 wt.% hydrogen desorption and reversible hydrogen absorption at a high temperature of 473 K. At this temperature, the first metallic C14 phases were fully hydrogenated and changed into their hydride counterparts. The structure demonstrated its structural flexibility for hydrogen storage by returning to a single C14 Laves phase upon dehydrogenation.

Because of their large number of interstitial sites, HEAs with a BCC structure are especially well-suited for storing hydrogen. According to Hu et al. (2021), for example, the Ti-Zr-V-Mo-Nb alloy has a theoretical hydrogen storage capacity of up to 2.65 wt.%, and when the hydrogen content surpasses 1.5 wt.%, it undergoes a phase transition from a BCC to an FCC structure [23]. Moreover, the Ti$_{35}$V$_{30}$Nb$_{10}$Cr$_{25}$ alloy recently set a record for BCC-structured HEAs with a storage capacity of 3.7 wt.%, underscoring the promise of BCC alloys for high-performance hydrogen storage applications [24]. These findings emphasize the potential of both BCC and Laves-phase HEAs to meet the growing demand for efficient and reversible hydrogen storage materials. In lightweight HEAs, achieving high hydrogen storage capacities remains challenging. The Mg-V-Al-Cr-Ni HEA, which is made via high-energy ball milling and has a single BCC structure, was investigated by Strozi et al. in 2021 [25]. Despite being lightweight, this alloy could not produce high-entropy hydrides and had a relatively poor hydrogen storage capacity of about 0.3 wt.%. This suggests that changes to the composition or synthesis methods may be required to increase the alloy hydrogen storage capacity. A similar study by Cardoso et al. (2021) looked at a Mg-Al-Ti-Fe-Ni HEA that was similarly made by high-energy ball milling. It showed a small hydrogen storage capacity (0.94 wt.%) but good kinetics, suggesting that it could still absorb and release hydrogen efficiently [26]. For real-world applications, the durability of hydrogen storage capacity under repeated cycles of hydrogen absorption and desorption is essential. A V$_{0.3}$Ti$_{0.3}$Cr$_{0.25}$Mn$_{0.1}$Nb$_{0.05}$ HEA that crystallised in a BCC phase with minimal secondary phases was studied by Liu et al. (2022) [27]. This alloy exhibited stable hydrogen desorption characteristics and a reversible hydrogen capacity of 1.78 wt.%. But over cycles, the capacity progressively decreased, mostly as a result of the room temperature incomplete desorption that produced stable hydrides. Even yet, after ten cycles, the alloy maintained 86% of its original capability, suggesting that structural modifications could lessen cyclic degradation. A Ti-Zr-Nb-Ta HEA hydrogenation behaviour was examined by Zhang et al. (2020), who found that several hydrides, including ε-ZrH2 and ε-TiH2, formed throughout a temperature range of 293 K to 493 K [28]. As the temperature rose, the maximum hydrogen capacity dropped from 1.67 wt.% to 1.25 wt.%, indicating a temperature-dependent phase stability. Despite the development of several hydride phases, the Ti-Zr-Nb-Ta HEA showed good hydrogen absorption kinetics and retained the majority of its capacity over cycles, indicating strong structural integrity. The CrFe-based Ti-Zr-Nb-Cr-Fe HEA, which combines the BCC and C14 Laves phases, was synthesised by ball milling and reported by Floriano et al. in 2021 [29]. This alloy demonstrated its efficacy at high temperatures, which facilitates both hydrogen absorption and desorption kinetics, with a reversible hydrogen storage capacity of 1.9 wt.% at 473 K. Computational methods that provide information on atomic-scale interactions in HEAs have recently validated the experimental results. $H_2$–Al interactions are weaker than $H_2$–Cr and $H_2$–Fe interactions, according to Li et al. (2024) analysis of hydrogen adsorption on Fe, Cr, and Al surfaces using density functional theory (DFT) [30]. According to this study, the alloying effects on $H_2$ dissociation differ according on the atomic composition, and Cr aggregation strengthens hydrogen binding. Future experimental designs will be guided by these insights, which advance our theoretical knowledge of hydrogen interactions in HEAs.

Based on the encouraging results with other HEAs, the Al-Cu-Fe-Ni-Cr HEA is being investigated for possible uses in hydrogen storage. This alloy has a single-phase BCC structure and was synthesized by high-energy ball milling. According to preliminary findings, cyclic stability and favourable hydrogen sorption behaviours are essential for long-term hydrogen storage. It is believed that the combination of these components balances the sorption kinetics and structural integrity, making it a viable option for additional study.

## 2. Materials and methods

### 2.1 Formation of single-phase Al-Cu-Fe-Ni-Cr high-entropy alloy:

The synthesis of the Al-Cu-Fe-Ni-Cr HEA began with high-purity elemental powders of aluminum, copper, iron, nickel, and chromium. These powders were sourced from reliable vendors such as Alfa Aesar for aluminum and chromium, Riedel-de Haen for iron, and Cerak Specialty Inorganics and Loba Chemie for nickel and copper. To ensure a uniform alloy, the elemental powders were combined in equal atomic percentages (20 at. % each) and mixed with hexane, which served as a process control agent (PCA). This facilitated the mechanical alloying process by preventing the powders from sticking together and improving the milling efficiency. The mechanical alloying was performed using an Attritor ball mill, also known as a Stirred ball mill or Szegvari Attritor [31]. The mill operated with a ball-to-powder ratio of 40:1 at a speed of 400 rpm. This particular ball mill is a vertical type with an agitator, designed to provide high wear resistance and consistent milling action. It features a 1.5-liter stainless steel container (mixing drum) equipped with a water jacket to control the temperature during the milling process, ensuring the alloy properties are not adversely affected by heat. Stainless steel balls, each with a diameter of 7 mm and a volume of 0.179503 cm³, were used as the grinding media. The consistent size and volume of the balls contributed to uniform energy transfer and effective mechanical alloying. After 40 hours of continuous milling, a BCC single-phase HEA with a lattice parameter of 0.289 nm was successfully synthesized. This structure was confirmed through X-ray diffraction and other characterization techniques. Following the synthesis, the hydrogenation and dehydrogenation properties of the HEA were studied. These properties are crucial for applications in hydrogen storage and other energy-related fields. The synthesized HEA demonstrated promising characteristics, paving the way for further research and potential applications in advanced materials science.

### 2.2 Characterization tools:

The phase of the HEA was evaluated using X-ray diffraction (XRD) with Cu K-alpha radiation (wavelength = 1.5406 Å), operating at 45 kV and 40 mA, and a scan step of 0.026°. Characterizations were performed using hybrid detector technology (PIXcel3D). The microscopic structure and chemical homogeneity of the Al-Cu-Fe-Ni-Cr HEA were examined with transmission electron microscopy (TEM) and energy-dispersive X-ray spectroscopy (EDX). TEM was performed at 200 kV using a TECNAI 20 G$^2$ FEA. The shape and topography of the nanocrystalline HEA were validated using a FESEM-Sigma300 with EDS, running at an accelerating voltage of 30 kV and a probe current of 20 nA, and utilizing a backscatter detector (BSD/HDBSD) at the Zeiss microscopic facility. X-ray photoelectron spectroscopy (XPS) was employed to analyze the electrical state and elemental composition of the alloy surface. XPS measurements were carried out under ultra-high vacuum (UHV) conditions at 25 W using the PHI5000 Versa Probe, with monochromatic AlKα radiation at an excitation energy of 1486.6 eV and a spot size of 100 micrometers. High-resolution core-level scans were obtained with a pass energy of 26 eV and a step size of 0.5 eV, while low-resolution survey scans were collected with a pass energy of 100 eV and a step size of 1 eV. Hydrogen sorption measurements were conducted using 400 mg of the sample, run through pressure composition isotherm (PCI) equipment provided by Advanced Material Corporation (Pittsburgh, USA).

The following formulas are used for calculating the thermodynamic parameters necessary for the synthesis of high entropy alloys (HEAs), such as mixing entropy (ΔSmix), mixing enthalpy (ΔHmix), atomic size difference (δ), valence electron concentration (VEC), and the Ω parameter [32]:

$$\Delta S_{mix} = -R \sum C_i \ln C_i \tag{1}$$

where R is a gas constant and Ci is molarity.

$$\Delta H_{mix} = \sum_{i=1, i \neq j}^{n} \Omega_{ij} C_i C_j \tag{2}$$

$$\Omega_{ij} = 4\Delta_{mix}^{AB}$$

where $\Delta_{mix}^{AB}$ is the mixing enthalpy of binary liquid AB alloys.

$$\delta = 100 \sqrt{\sum_{i=1}^{n} C_i (1 - \frac{r_i}{\bar{r}_i})^2} \quad, \bar{r} = \sum_{i=1}^{n} C_i r_i, \tag{3}$$

where ri and Ci stand for atomic radius and atomic percentage, respectively.

$$VEC = \sum_{i=1}^{n} C_i (VEC)_i, \tag{4}$$

where (VEC)$_i$ stands for the valence electron concentration of the i-th component.

$$\Omega = \frac{T_m \Delta S_{mix}}{\Delta H_{mix}} \tag{5}$$

where $T_m$ is the melting temperature.

### 3. Results and discussion:
### 3.1 Overview of Structural, Microstructural, and Surface Morphology Analysis:

As-synthesized HEA, the dehydrogenated HEA following several cycles of hydrogenation and dehydrogenation, and the initial powder are all structurally characterized using X-ray diffraction (XRD) techniques in Figure 1. The first hand-mixed powder XRD pattern in Figure 1(a) displays a prominent Al (111) peak, suggesting that Al is the phase with the highest intensity. A single-phase BCC structure with a lattice parameter of 0.212 nm, attributed to the space group Im-3m, is revealed by the XRD pattern of the as-synthesized HEA, as shown in Figure 1(b). On the other hand, the XRD pattern of the dehydrogenated Al-Cu-Fe-Ni-Cr HEA following many cycles of hydrogenation and dehydrogenation is displayed in Figure 1(c). The Body-Centered Tetragonal (BCT) structure is used in this instance, and the lattice parameters are a = b = 0.232 nm and c = 0.3655 nm. The coexistence of two BCC phases, a single BCT phase, and the coexistence of BCC and BCT phases were among the refinement models that were examined. A single BCT phase produced the finest match. This outcome is consistent with earlier research showing that completely hydride BCC alloys frequently adopt the FCC structure, while BCT structures are usually seen as monohydride phases of BCC alloys. But according to Sahlberg et al. (2016), the HEA Ti-V-Zr-Hf-Nb dihydride phase also takes on a BCT structure [33]. They ascribed this to lattice strain in the HEA matrix, which permits the occupation of both tetrahedral and octahedral sites, hence supporting a high hydrogen concentration. Regardless of particle size, temperature, or pressure, Sleiman et al. (2021) also noted that hydrogenation changed the BCC structure to BCT [34]. Many transition-metal-hydrogen systems have been found to exhibit a distorted BCC phase at intermediate hydrogen content (BCC → distorted BCC, or BCT). The present results align with the findings of Sleiman et al. and Sahlberg et al. Le Bail profile fitting was used using the JANA 2006 program to improve the XRD patterns of the as-synthesised and dehydrogenated HEAs. This sophisticated fitting method allowed for a thorough examination of the intricate diffraction patterns, as seen in Figure 2 (a-b). The samples phase purity and structural integrity are supported by the refinement data, which includes the unit cell parameters given in Table 1. Both the as-synthesised and dehydrogenated Al-Cu-Fe-Ni-Cr HEA samples EDS spectra, bright-field TEM pictures, and selected area electron diffraction (SAED) patterns are shown in Figure 3. A bright-field TEM picture of the as-synthesised HEA is shown in Figure 3(a), and the corresponding SAED pattern, indexed with the (110) plane, is shown in Figure 3(d), which confirms the XRD results. A nanoscale crystallinity in the sample is indicated by the SAED pattern patchy ring shape, which is typical of nanocrystalline materials. Figure 3(b) displays a bright-field TEM picture of the dehydrogenated Al-Cu-Fe-Ni-Cr HEA, demonstrating both porosity and nanoparticles. Porosity may increase hydrogen storage capacity because it indicates structural changes or the creation of voids that are probably the result of the hydrogenation and dehydrogenation processes. Similar to the as-synthesised sample, the SAED pattern of the dehydrogenated HEA, displayed in Figure 3(e), also displays a spoty ring structure, demonstrating the nanocrystalline nature and retention of a crystalline structure after dehydrogenation. The EDS spectra for the as-synthesised and dehydrogenated materials are shown in Figures 3(c) and 3(f), respectively. Both samples expected elemental composition is confirmed by the EDS analysis, which also shows that there are no notable contaminants present. The HEA retains its structural integrity and purity throughout the hydrogenation and dehydrogenation cycles, as evidenced by the elemental composition constancy.

The surface morphology and elemental content of the as-synthesized Al-Cu-Fe-Ni-Cr HEA and the dehydrogenated HEA following several hydrogenation-dehydrogenation cycles were examined using EDS in combination with SEM, as shown in Figure 4(a-d). A flaky texture can be seen in the SEM picture of the as-synthesized HEA (Figure 4(a)). Particle fracturing, cold welding, mechanical deformation, and surface roughening are all caused by the mechanical impacts and frictional forces that occur during the ball milling process, which is why the surface morphology is uneven and rough. Following hydrogen cycling, the dehydrogenated HEA exhibits notable morphological changes. The surface fissures seen in the SEM picture (Figure 4(b)) are most likely the result of lattice expansion during the hydrogenation process. These findings support the structural alterations brought about by hydrogen cycling and are in agreement with the XRD data. Using EDS analysis, the elemental makeup of the dehydrogenated and as-synthesised HEA was verified. The EDS spectra, which are displayed in Figures 4(c) and 4(d), demonstrate the purity of the synthesised alloy by confirming the lack of impurity elements.

### 3.2 X-ray photoelectron spectroscopy (XPS) analysis of Al-Cu-Fe-Ni-Cr HEA:

To examine the fundamental composition and chemical structure of the Al-Cu-Fe-Ni-Cr HEA, we performed XPS analysis on both the as-synthesized HEA and the dehydrogenated sample following several cycles of hydrogenation and dehydrogenation. Important details about the electronic setup of the HEA under investigation were revealed by this thorough analysis [35]. Both the as-synthesized and dehydrogenated HEA samples deconvoluted XPS spectra, with high-resolution and low-resolution survey scans, are shown in Figures 5 and 6 to show the variations in electronic states under each scenario. The two peaks in the Al XPS spectra for the as-synthesized HEA are located at 75.5 eV ($2P_{1/2}$) and 72.5 eV ($2P_{3/2}$), which represent the +2 and +3 oxidation states, respectively [36]. These peaks, which were still indicative of the +2 and +3 oxidation states, moved to 77 eV ($2P_{1/2}$) and 74 eV ($2P_{3/2}$) following dehydrogenation. Lattice strain and defect development during cycles of hydrogenation and dehydrogenation are responsible for this change in binding energy. Cu in the as-synthesised HEA showed satellite peaks at 960.5 eV and 940.5 eV, as well as peaks at 950.5 eV ($2P_{1/2}$) and 931.5 eV ($2P_{3/2}$), which correspond to the +2 and +3 oxidation states [37]. While keeping the same oxidation states, these peaks moved to 953 eV ($2P_{1/2}$) and 933 eV ($2P_{3/2}$) in the dehydrogenated HEA, while satellite peaks moved to 961.5 eV and 941.5 eV. Consistent with +2 and +3 oxidation states, Fe in the as-synthesised HEA showed peaks at 723 eV ($2P_{1/2}$) and 710 eV ($2P_{3/2}$) in addition to a satellite peak at 715 eV [38]. These peaks moved to 724 eV ($2P_{1/2}$) and 711 eV ($2P_{3/2}$) in the dehydrogenated HEA, and a new satellite peak was detected at 717 eV, suggesting a shift in the electronic environment. In the same way, Ni in the as-synthesised HEA had satellite peaks at 878 eV and 860 Ev, along with peaks at 871 eV ($2P_{1/2}$) and 854 eV ($2P_{3/2}$) which corresponded to +2 and +3 oxidation states [39]. The dehydrogenated sample had satellite peaks at 880 eV and 861 eV, along with slightly shifted peaks at 873 eV ($2P_{1/2}$) and 855 eV ($2P_{3/2}$), indicating minor oxidation state changes. Peaks for Cr were found in the as-synthesised HEA at 575 eV ($2P_{1/2}$) and 569 eV ($2P_{3/2}$), which correspond to +2 and +3 oxidation states, respectively [40-41]. Without any satellite peaks, these peaks moved slightly to 586 eV ($2P_{1/2}$) and 579 eV ($2P_{3/2}$) during dehydrogenation. The lattice strain, defects, and changes in the electronic environment brought on by the hydrogenation and dehydrogenation cycles are probably the causes of these shifts in binding energies among the various elements in the HEA.

### 3.3 Study of re/dehydrogenation kinetics and thermodynamics with cyclic Stability in Al-Cu-Fe-Ni-Cr HEA:

The Al-Cu-Fe-Ni-Cr HEA hydrogen absorption and desorption behaviour was thoroughly examined, as shown in Figures 7(a) and 7(b). To begin the analysis, the HEA was put through ten activation cycles at room temperature with a hydrogen pressure of 100 atm. Following that, measurements of the absorption and desorption capabilities were made at room temperature (RT), 100°C, 200°C, and 300°C. Hydrogen pressures were kept at 50 atm during the absorption phase, while desorption was aided by 0.1 atm. At room temperature and 100°C, the Al-Cu-Fe-Ni-Cr HEA demonstrated exceptional hydrogen absorption efficiency, obtaining hydrogen uptakes of 1.7 wt.% and 1.8 wt.% in 4 minutes, respectively. These figures highlight the HEA capacity to store hydrogen efficiently at comparatively low temperatures. The alloy hydrogen absorption increased even more when the temperature was raised to 200°C and 300°C, reaching 1.9 wt.% and 2.1 wt.% in under 5 minutes. This modest rise in absorption capacity at higher temperatures raises the possibility that higher temperatures could lower kinetic barriers and promote better diffusion of hydrogen into the alloy matrix. The alloy maximum absorption rate was reached at 300°C, demonstrating both its improved kinetics efficiency for hydrogen storage applications and its potential for enhanced hydrogen uptake at high temperatures. Figure 7(b) shows the desorption behaviour of the Al-Cu-Fe-Ni-Cr HEA at various temperatures. The alloy emitted 0.4 wt.% and 0.5 wt.% of hydrogen at room temperature and 100°C, respectively, suggesting a moderate rate of desorption under these circumstances. A minor improvement in desorption was noted at 200°C, where 0.9 wt.% hydrogen was released. The alloy highest release of about 1.6 wt.% in about 12 minutes was the most notable desorption that took place at 300°C, highlighting the HEA improved desorption kinetics at higher temperatures. All things considered, the results show that the Al-Cu-Fe-Ni-Cr HEA exhibits effective hydrogen cycling properties, with especially encouraging performance at 300°C, making it a solid contender for applications needing quick hydrogen absorption and release. Figure 8 illustrates the temperature-programmed desorption (TPD) investigation that provided additional understanding of the hydrogen desorption properties. Desorption stayed slow up to about 200°C, according to the TPD study, which was carried out at a heating rate of 7°C/min up to 450°C. Desorption kinetics showed a noticeable rise at this point, and the release rate kept speeding up as the temperature rose, approaching near-complete desorption by about 350°C. The HEA potential for significant hydrogen release under regulated heating settings was confirmed by the recording of a peak desorption capacity of 1.9 wt.%. Important information about the Al-Cu-Fe-Ni-Cr HEA hydrogen storage behaviour is shown in Figures 9(a) and 9(b). During absorption and desorption processes at room temperature, 100°C, 200°C, and 300°C, the Pressure-Composition-Isotherms (PCI), shown in Figure 9(a), were recorded. Measurements were conducted at a hydrogen pressure of 15 atm. Stable hydrogen absorption and desorption processes are indicated by the PCI curves low equilibrium pressures for both absorption and desorption, which stay below 7 atm at all temperatures. The lowest storage capacity was recorded at RT (1.7 wt.%), while the maximum capacity was recorded at 300°C (2.1 wt.%). The alloy favourable absorption and desorption behaviour is further demonstrated by the flat pressure plateaus seen on the isotherms, demonstrating its appropriateness for real-world hydrogen storage applications. Figure 9(b) shows the cyclic stability of the Al-Cu-Fe-Ni-Cr HEA over 25 cycles of hydrogen absorption and desorption. For real-world applications where reversibility is crucial, this stability is crucial. The alloy storage capacity only marginally dropped from 2.1 wt.% to 1.9 wt.% over the course of 25 cycles, demonstrating its exceptional cycle stability and confirming its potential as a workable material for hydrogen storage devices.

The dynamic evolution of the structure during repeated hydrogen cycling is reflected in the Al-Cu-Fe-Ni-Cr HEA BCC-to-BCT transition. Effective hydrogen absorption and desorption with strong cycle stability are made possible by the BCC phase open lattice structure and uniform interstitial sites in its as-synthesised condition [42]. Nevertheless, the transition to the BCT phase is brought on by accumulated lattice strain and anisotropic stress with repeated cycles of hydrogenation and dehydrogenation. Both cyclic stability and hydrogen storage capacity are impacted by this structural change, which causes lattice anisotropy, decreases symmetry, and increases internal strain. In this HEA, hydrogen storage starts at the alloy surface, where catalytic reactions cause hydrogen molecules ($H_2$) to split into atomic hydrogen (H) [43]. Because of its remarkable catalytic qualities, Ni is essential at this stage, greatly accelerating the kinetics of hydrogen dissociation and absorption [44]. By promoting the development of distortion-tolerant phases, Ni also helps the alloy remain stable throughout cycling. Ni also promotes quicker and more effective hydrogen release by lowering the activation energy needed for hydrogen desorption. By offering more locations for hydrogen dissociation and absorption, Fe enhances Ni catalytic activity. Fe also increases the alloy mechanical strength, which helps it withstand structural deterioration during several cycles of hydrogenation and dehydrogenation [45]. The HEA high configurational entropy encourages hydride stabilisation, which is essential for boosting hydrogen storage capacity. Hydrogen atoms diffuse into the lattice during hydrogen absorption, taking up interstitial spaces and combining with other metallic elements to produce hydrides [46]. The opposite happens during desorption, when hydrogen atoms return to the lattice after exiting the hydride phase. Eventually, these atoms desorb from the alloy surface and recombine into $H_2$ molecules. Cu and Al promote the storage of hydrogen [47]. Al helps to generate stable hydrides, which increases specific hydrogen capacity [48]. Additionally, the alloy overall hydrogen storage efficiency is enhanced by its lightweight nature. Cu, on the other hand, stabilises the microstructure and increases the alloy endurance over time. Cu increases heat conductivity, which makes it easier for hydrogen to be released efficiently, even though it does not easily form hydrides [49]. By strengthening corrosion resistance and shielding the alloy from oxidative deterioration during cycling, Cr further assists the system [50]. By causing lattice distortions that produce more interstitial sites for hydrogen storage, Cr boosts the capacity of hydrogen storage [51]. A compromise between hydride stability and effective absorption/desorption without excessive hysteresis is guaranteed by its moderate affinity for hydrogen. Following several cycles of hydrogenation and dehydrogenation, XPS data show a rise in the principal elements binding energy in the dehydrogenated Al-Cu-Fe-Ni-Cr HEA. This rise is ascribed to variations in the alloy electrical and chemical environments as well as microstructural changes. Significant lattice distortions, including dislocations, vacancies, and grain boundary movements, are introduced by repeated cycles of hydrogen absorption and desorption. Higher binding energies are the outcome of these faults because they put constituent elements in a more strained and energetically unfavourable environment. The shifting interactions within the alloy microstructure are reflected in the variations in binding energies [52]. Additionally, some elements surface segregation is encouraged by hydrogen cycling. For example, during cycling, Ni prefers to move to the surface because of its catalytic nature [53]. Because of variations in the coordination environment relative to the bulk, this surface segregation modifies the local composition and increases the binding energy of surface components. SEM-EDS analysis, which shows alterations in the surface morphology and elemental distribution of the HEA, provides evidence of this surface segregation. The alloy dynamic reorganisation and the change from BCC to BCT demonstrate the intricate interactions between structural, chemical, and catalytic elements that control hydrogen storage and cyclic stability in the Al-Cu-Fe-Ni-Cr HEA.

4. **Conclusions:**

The Al-Cu-Fe-Ni-Cr HEA was successfully synthesised in this study utilising high-energy attritor ball milling, yielding a single-phase nanocrystalline alloy with a BCC structure and a lattice parameter of 0.289 nm. With its quick rates of absorption and desorption, this HEA showed remarkable hydrogen storage properties. In just 3 minutes, the alloy absorbed about 2.1 wt.% hydrogen at 300°C and 50 atm of hydrogen pressure. At 300°C and 0.1 atm, desorption was likewise effective, releasing almost 1.6 wt.% hydrogen in 6 minutes. The Al-Cu-Fe-Ni-Cr HEA is a very promising material for hydrogen storage applications because of its quick and effective hydrogen sorption capabilities. Furthermore, it demonstrated excellent cyclic stability, sustaining a negligible capacity deterioration of 0.2 wt.% across 25 cycles of hydrogenation and dehydrogenation. The material long-term stability and dependability, which are essential for sustainable hydrogen storage, are shown by this slight capacity loss. These encouraging results add to HEAs expanding potential to advance solid-state hydrogen storage technologies.

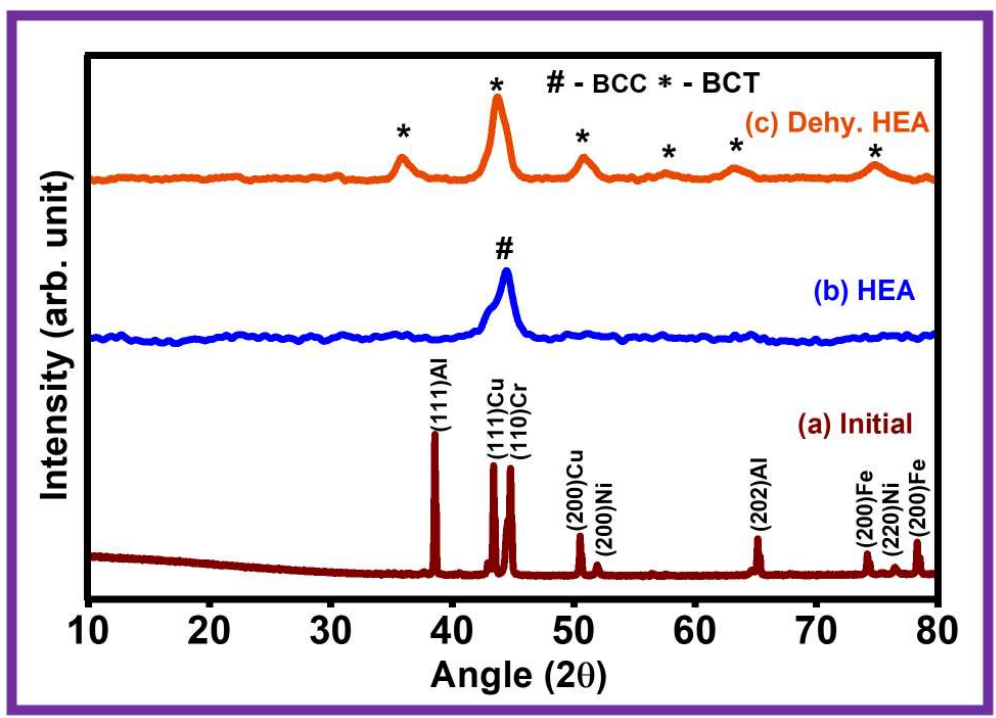

**Fig. 1:** X-ray diffraction (XRD) patterns showcasing the evolution of the Al-Cu-Fe-Ni-Cr high entropy alloy (HEA) at various stages: (a) the HEA diffraction pattern when it was first powdered, (b) the diffraction pattern following synthesis, and (c) the diffraction pattern of the dehydrogenated HEA following cycles of rehydrogenation and dehydrogenation.

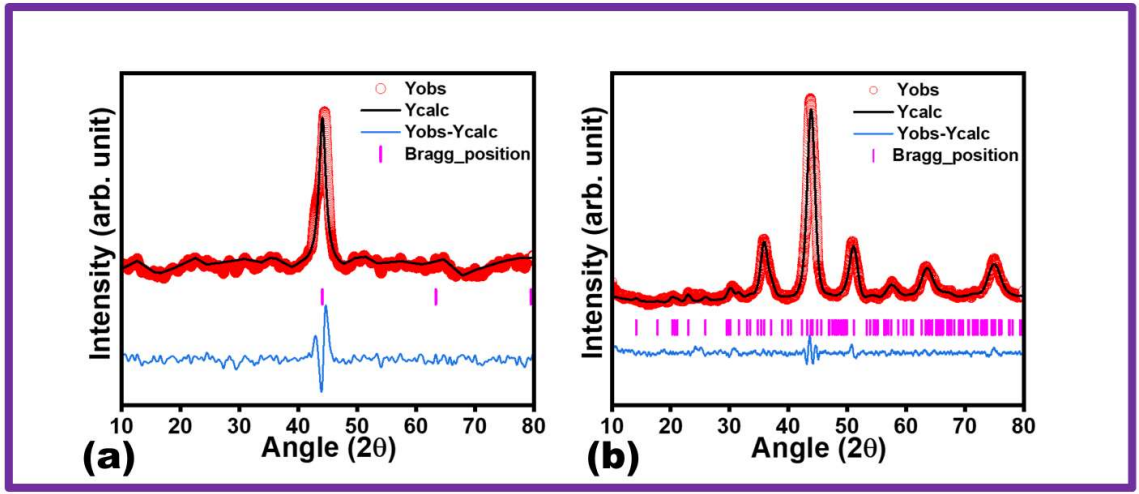

**Fig 2:** Le Bail profile fit parameters showing the Al-Cu-Fe-Ni-Cr high entropy alloy (HEA) phase characterisation: (a) Al-Cu-Fe-Ni-Cr HEA as-synthesised, and (b) dehydrogenated Al-Cu-Fe-Ni-Cr HEA.

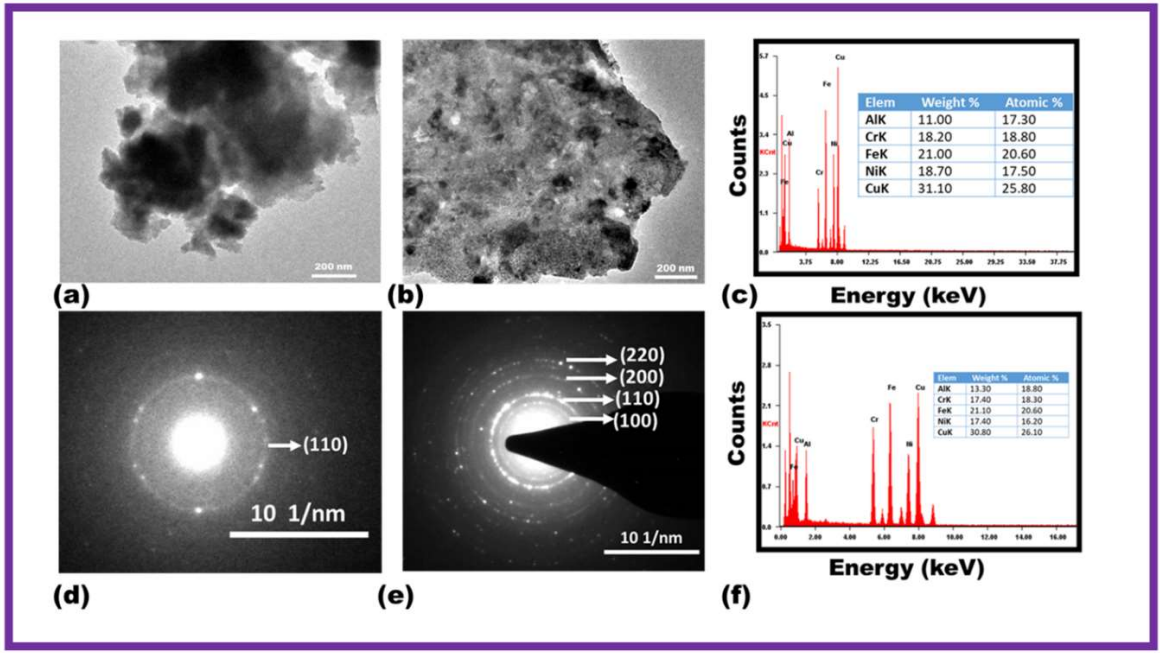

**Fig 3:** Microscopic and spectroscopic analysis of the Al-Cu-Fe-Ni-Cr high-entropy alloy (HEA) includes: (a) TEM micrograph of the as-synthesized HEA, (b) TEM micrograph of the dehydrogenated HEA showing discernible nanoparticles in bright field imaging, (c) EDS spectra of the as-synthesized HEA, (d) corresponding SAED pattern of the as-synthesized HEA, (e) corresponding SAED pattern of the dehydrogenated HEA, and (f) EDS spectra of the dehydrogenated HEA.

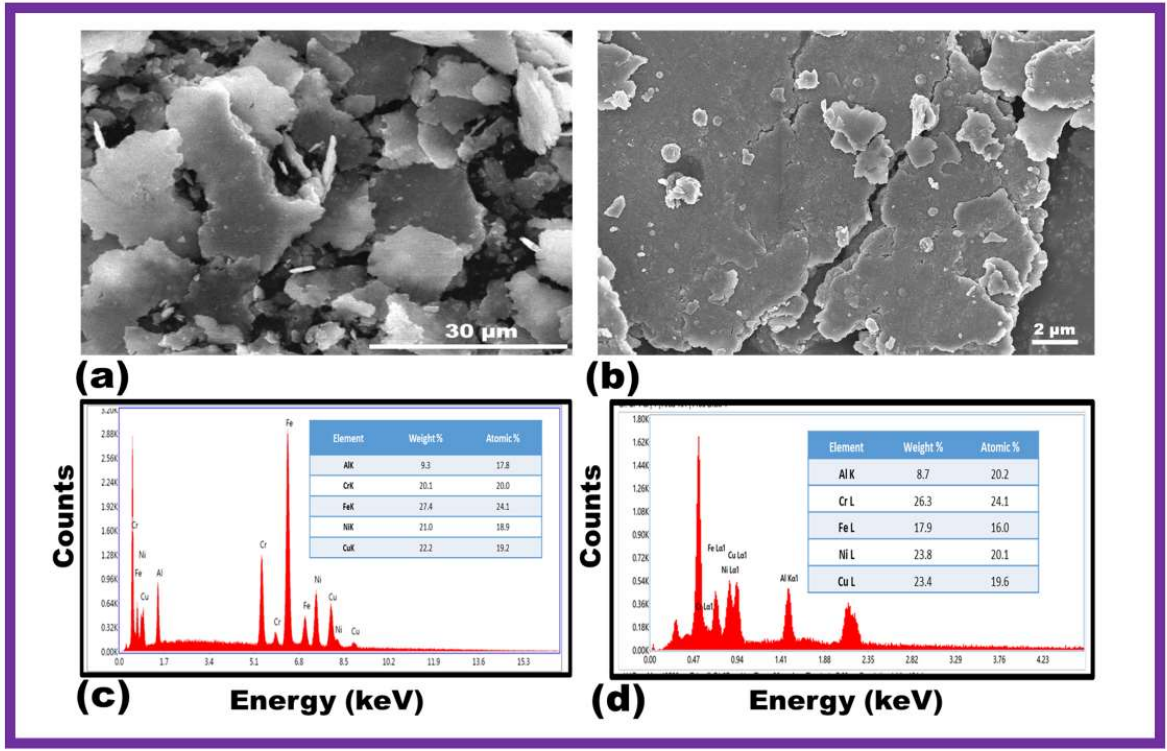

**Fig 4:** Microstructural and compositional analysis of Al-Cu-Fe-Ni-Cr high entropy alloy (HEA): (a) SEM micrograph of the HEA as synthesised, (b) SEM micrograph of the dehydrogenated HEA following rehydrogenation-dehydrogenation cycling, displaying cracks caused by lattice expansion during hydrogenation, (c) EDS spectra featuring elemental composition analysis of the as-synthesised HEA, and (d) EDS spectra with elemental composition analysis of the dehydrogenated HEA following rehydrogenation-dehydrogenation cycling.

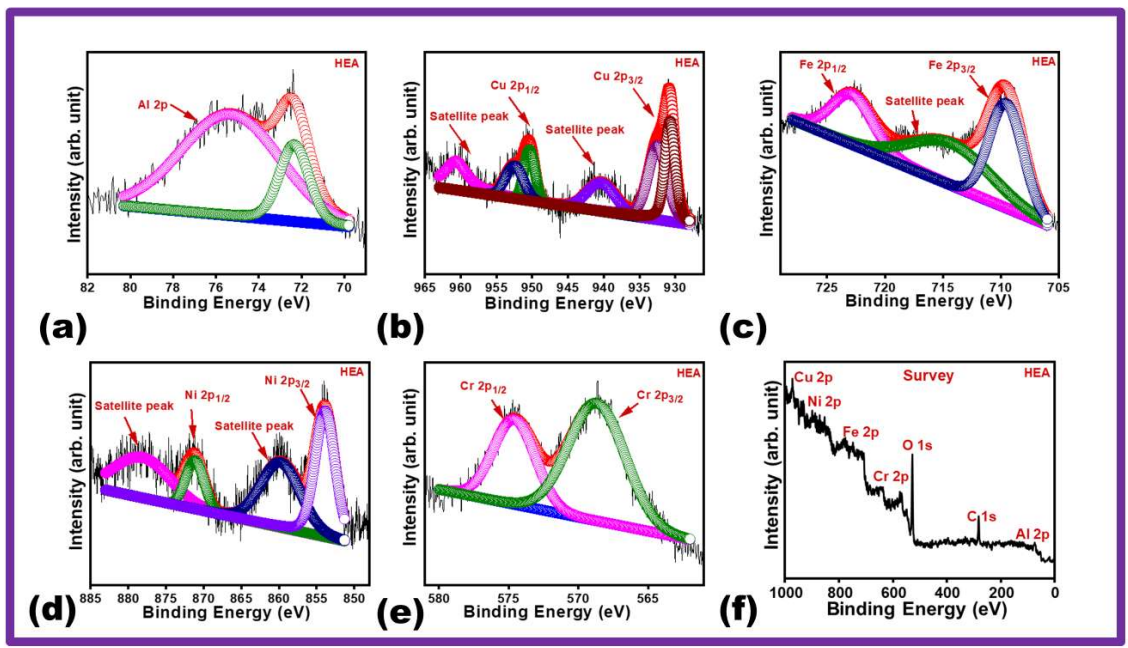

**Fig 5:** As-synthesised Al-Cu-Fe-Ni-Cr high entropy alloy (HEA) analysed using X-ray photoelectron spectroscopy (XPS): (a) high-resolution spectrum of Al, (b) high-resolution spectrum of Cu, (c) high-resolution spectrum of Fe, (d) high-resolution spectrum of Ni, (e) high-resolution spectrum of Cr, and (f) low-resolution survey scan revealing the elemental composition.

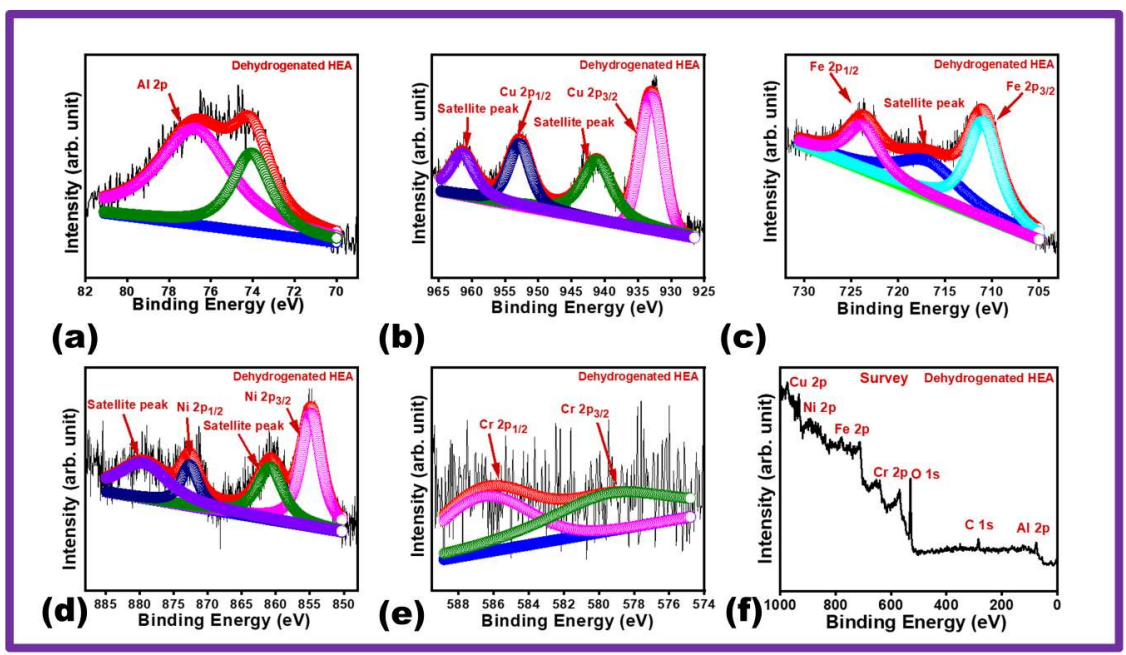

**Fig 6:** Analysis of the dehydrogenated Al-Cu-Fe-Ni-Cr high entropy alloy (HEA) using X-ray Photoelectron Spectroscopy (XPS) following rehydrogenation-dehydrogenation cycling: Spectra of the 2p areas in high resolution for (a) Al, (b) Cu, (c) Fe, (d) Ni, and (e) Cr, as well as (f) a survey scan in low resolution.

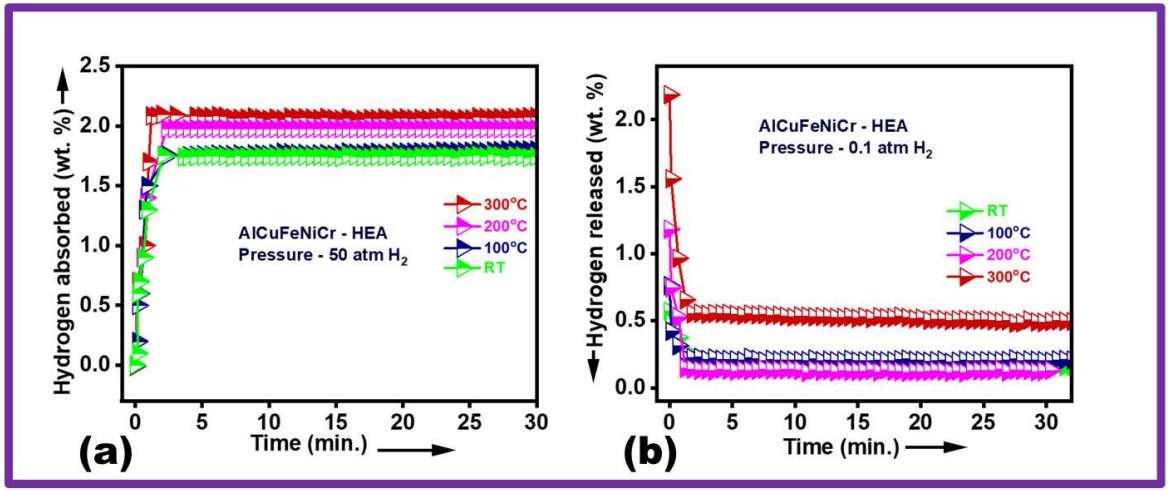

**Fig 7:** Hydrogen sorption behaviour of the as-synthesised Al-Cu-Fe-Ni-Cr high entropy alloy (HEA): (a) Rehydrogenation curves under 50 atm $H_2$ pressure at room temperature (RT), 100°C, 200°C, and 300°C, and (b) dehydrogenation curves under 0.1 atm $H_2$ pressure at RT, 100°C, 200°C, and 300°C.

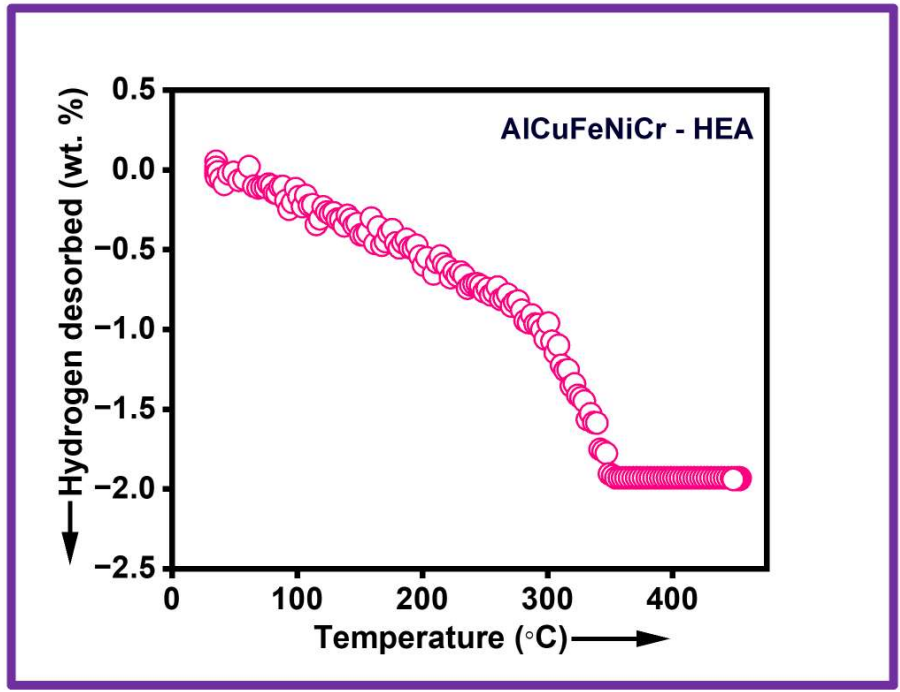

**Fig 8:** Temperature Programmed Desorption (TPD) curves for the hydrogenated Al-Cu-Fe-Ni-Cr high entropy alloy.

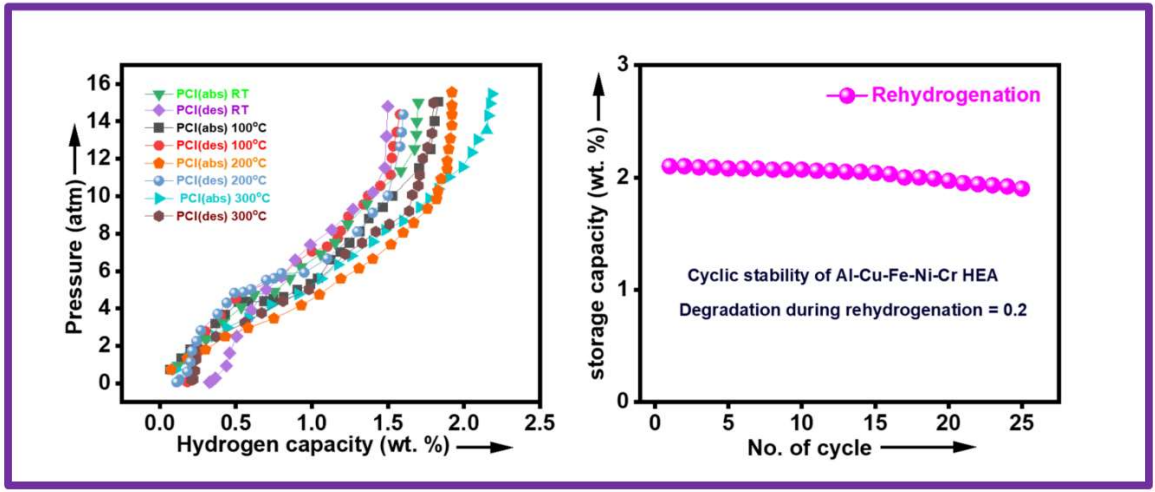

**Fig 9:** hydrogen storage performance of Al-Cu-Fe-Ni-Cr high entropy alloy (HEA): (a) Pressure-composition isotherms (PCI) for hydrogen absorption and desorption at three distinct temperatures, and (b) cyclic stability curves showing the as-synthesised HEA rehydrogenation capacity over 25 cycles at 300°C under 50 atm hydrogen pressure.